\documentclass[%
superscriptaddress,
reprint,
showpacs,
aps,
pra,
]{revtex4-1}
\usepackage{graphicx}
\usepackage{dcolumn}
\usepackage{bm}
\usepackage{braket}
\usepackage{amsthm}
\usepackage{epstopdf}
\usepackage{amssymb}
\usepackage{amsmath}
\usepackage{bbold}
\usepackage{hyperref}
\usepackage{comment}
\usepackage{xcolor}
\usepackage{ulem}

\newcommand{\tr}[1]{\text{Tr}[{#1}]}

\theoremstyle{definition}

\begin{document}
\preprint{APS/123-QED}
\title{Optimal measurements for quantum fidelity between Gaussian states and its relevance to quantum metrology}
\author{Changhun Oh}%
\email{v55ohv@snu.ac.kr}
\affiliation{Center for Macroscopic Quantum Control, Department of Physics and Astronomy, Seoul National University, Seoul 08826, Korea}
\author{Changhyoup Lee}
\email{changhyoup.lee@gmail.com}
\affiliation{Institute of Theoretical Solid State Physics, Karlsruhe Institute of Technology, 76131 Karlsruhe, Germany}
\author{Leonardo Banchi}
\affiliation{QOLS, Blackett Laboratory, Imperial College London, London SW7 2AZ, United Kingdom}
\author{Su-Yong Lee}
\affiliation{School of Computational Sciences, Korea Institute for Advanced Study, Hoegi-ro 85, Dongdaemun-gu, Seoul 02455, Korea}
\author{Carsten Rockstuhl}
\affiliation{Institute of Theoretical Solid State Physics, Karlsruhe Institute of Technology, 76131 Karlsruhe, Germany}
\affiliation{Institute of Nanotechnology, Karlsruhe Institute of Technology, 76021 Karlsruhe, Germany}
\author{Hyunseok Jeong}
\affiliation{Center for Macroscopic Quantum Control, Department of Physics and Astronomy, Seoul National University, Seoul 08826, Korea}
\date{\today}
\begin{abstract}
Quantum fidelity is a measure to quantify the closeness between two quantum states. In an operational sense, it is defined as the minimal overlap between the probability distributions of measurement outcomes and the minimum is taken over all possible positive-operator valued measures~(POVMs). Quantum fidelity has been investigated in various scientific fields, but the identification of associated optimal measurements has often been overlooked despite its great importance both for fundamental interest and practical purposes.  We find here the optimal POVMs for quantum fidelity between multimode Gaussian states in a closed analytical form. Our general finding is applied for selected single-mode Gaussian states of particular interest and we identify three types of optimal measurements: an excitation-number-resolving detection, a projection onto the eigenbasis of operator $\hat{x}\hat{p}+\hat{p}\hat{x}$, and a quadrature variable detection, each of which corresponds to distinct types of single-mode Gaussian states. We also show the equivalence between optimal measurements for quantum fidelity and those for quantum parameter estimation when two arbitrary states are infinitesimally close. It is applied for simplifying the derivations of quantum Fisher information and the associated optimal measurements, exemplified by displacement, phase, squeezing, and loss parameter estimation using Gaussian states.
\end{abstract}

\maketitle
\section{introduction}
Quantification of the similarity between quantum states is of the utmost importance in quantum information processing such as quantum error correction and quantum communication~\cite{nielsen2000, wilde2017, weedbrook2012, braunstein2005}. There are various measures of the closeness between two quantum states such as trace distance~\cite{helstrom1976}, quantum Chernoff bound~\cite{audenaert2007, audenaert2008}, and quantum relative entropy~\cite{vedral2002}.
Among the diverse measures, one of the most common measures is quantum fidelity~\cite{uhlmann1976}. Theoretically, it is defined as the minimal overlap of the probability distributions obtained by an optimal positive-operator valued measure~(POVM) performed on two states. It has also been widely employed to verify how close actual states are to target states in experiments~\cite{leibfried2004, lu2007, ourjoumtsev2007}, practically assessing quantum information processing protocols such as quantum teleportation~\cite{bennett1993, bouwmeester1997, braunstein1998, furusawa1998} and quantum cloning~\cite{buzek1996, lindblad2000, cerf2000, braunstein2001, fiurasek2001}. It has been known that the quantum fidelity not only plays a crucial role in quantum parameter estimation~\cite{helstrom1976, braunstein1994}, but also sets a bound for quantum hypothesis testing \cite{helstrom1976, fuchs1999} and, particularly, quantum Chernoff bound \cite{audenaert2007, audenaert2008}.

\begin{figure}[b]
\centering
\includegraphics[width=0.47\textwidth]{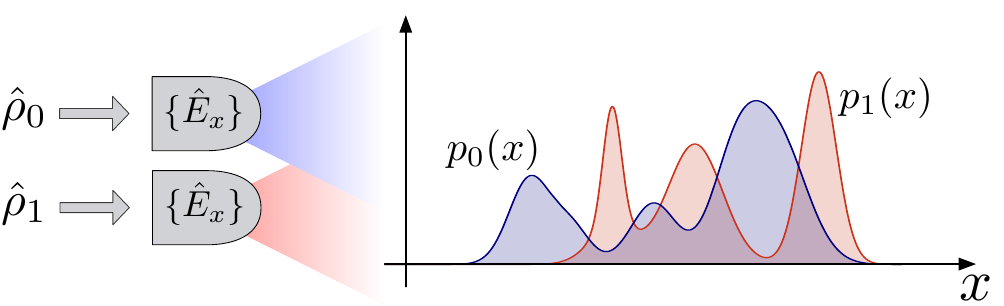}
\caption{Quantum fidelity between two states $\hat{\rho}_{0}$ and $\hat{\rho}_{1}$ can be measured by the minimal overlap between the probability distributions $p_0(x)$ and $p_1(x)$, where the measurement outcomes $x$ are obtained by an optimally chosen POVM $\{ \hat{E}_{x}\}$.}
\label{setting}
\end{figure}

One useful platform for quantum information processing is continuous-variable systems, such as optical fields with indefinite photon numbers~\cite{weedbrook2012}. In particular, bosonic Gaussian states are practical resources because they are relatively less demanding to generate and manipulate in experiments~\cite{ferraro2005, wang2007, weedbrook2012, adesso2014, serafini2017}.
Due to the importance of quantum fidelity between Gaussian states, there have been numerous attempts to find an analytical formula between constrained Gaussian states~\cite{twamley1996, nha2005, olivares2006, scutaru1998, marian2012, marian2003, marian2008, spedalieri2013, paraoanu2000, banchi2015}, but only recently arbitrary Gaussian states have been implemented in a computable analytical formula of quantum fidelity~\cite{banchi2015}.
The quantum fidelity can be obtained with the optimal POVM, but the optimal measurement setting achieving quantum fidelity between Gaussian states has not yet been found, although a general method of finding an optimal measurement for two given quantum states is known~\cite{fuchs1995}. Furthermore, an explicit relation between quantum fidelity and quantum Fisher information, found in Ref.~\onlinecite{braunstein1994}, raises a further intriguing question on the relevance of optimal measurements for quantum fidelity to those required for optimal quantum metrology.

In this work, we find the optimal POVMs, in a closed analytical form, enabling one to measure quantum fidelity between two multimode Gaussian states. Such general form of optimal POVMs allows us to classify optimal measurements for quantum fidelity between two single-mode Gaussian states of particular interest. 
In addition, we demonstrate the equivalence between optimal measurements for quantum fidelity and those for quantum Fisher information, upon which we discuss quantum parameter estimation in the context of single-mode Gaussian metrology~\cite{monras2007,pinel2013,safranek2016,oh2018}, such as displacement, phase, squeezing, and loss parameter estimation.

\textit{Preliminaries.}
Consider that a measurement described by a POVM $\{\hat{E}_x\}$ satisfying $\int dx\hat{E}_x=\mathbb{1}$ and $\hat{E}_x\geq0$ is performed on two states $\hat{\rho}_{0}$ and $\hat{\rho}_{1}$, yielding the probability distributions for outcomes $x$, written by $p_{j}(x)=\text{Tr}[\hat{\rho}_{j}\hat{E}_x]$ for $j=0,1$, as shown in Fig. 1.
One notable measure of statistical distinguishability of two probability distributions is the Bhattacharyya coefficient~\cite{fuchs1995, bhattacharyya1943, wilde2017}, written as
\begin{align*}
BC(p_{0},p_{1})=\bigg(\int \sqrt{p_0(x)p_1(x)}dx\bigg)^2.
\end{align*}
This quantity takes the maximum value of 1 if and only if two given probability distributions are equivalent, i.e., $p_0(x)=p_1(x)$ for all possible outcomes $x$.
This notion of distinguishability has been extended to the quantum regime by minimizing $BC(p_{0},p_{1})$ over all possible POVMs \{$\hat{E}_x$\}.
The quantum fidelity is thus defined as
\begin{align} \label{deffid}
F(\hat{\rho}_0,\hat{\rho}_1)=\min_{\{\hat{E}_x\}}BC(p_{0},p_{1}),
\end{align}
which further reduces to a known form as \cite{uhlmann1976}
\begin{align*}
F(\hat{\rho}_0,\hat{\rho}_1)=\left(\text{Tr}\sqrt{\hat{\rho}_1^{1/2}\hat{\rho}_0\hat{\rho}_1^{1/2}}\right)^2.
\end{align*}

From the definition of quantum fidelity in Eq.~\eqref{deffid}, it is obvious that finding the optimal POVM is crucial to maximally distinguish two given quantum states.
It has been found that the optimal measurements have to satisfy
\begin{align}
\hat{E}_{x}^{1/2}(\hat{\rho}_1^{1/2}-\mu_x\hat{\rho}_0^{1/2}\hat{W}^\dagger)&=0, \label{fidc1} \\
\text{Tr}(\hat{W}\hat{\rho}_0^{1/2}\hat{E}_{x}\hat{\rho}_1^{1/2})&\in\mathbb{R}, \label{fidc2}
\end{align}
where $\hat{W}$ is a unitary operator satisfying $\hat{W}\hat{\rho}_0^{1/2}\hat{\rho}_1^{1/2}=\sqrt{\hat{\rho}_1^{1/2}\hat{\rho}_0\hat{\rho}_1^{1/2}}$ and $\mu_x$ is a constant~\cite{fuchs1995}.
In the case of full-rank states $\hat{\rho}_0$ and $\hat{\rho}_1$, the optimal measurement $\{\hat{E}_x\}$ is unique and consists of projections onto the eigenbasis of a Hermitian operator written by
\begin{align} \label{meq}
\hat{M}(\hat{\rho}_0,\hat{\rho}_1)=\hat{\rho}_1^{-1/2}\sqrt{\hat{\rho}_1^{1/2}\hat{\rho}_0\hat{\rho}_1^{1/2}}\hat{\rho}_1^{-1/2}.
\end{align}
Thus, simplifying the operator $\hat{M}$ to find its eigenbasis is the central task to identify optimal measurements.
In addition, we note a simple, but highly useful property of the operator $\hat{M}$,
\begin{align}\label{Mproperty}
\hat{M}(\hat{U}\hat{\rho}_0\hat{U}^\dagger,\hat{U}\hat{\rho}_1\hat{U}^\dagger)=\hat{U}\hat{M}(\hat{\rho}_0,\hat{\rho}_1)\hat{U}^\dagger,
\end{align}
where $\hat{U}$ is a unitary operator.

\section{Optimal measurements for multi-mode Gaussian states}
Consider $n$ bosonic modes described by quadrature operators $\hat{Q}\equiv(\hat{x}_1,\hat{p}_1,\hat{x}_2,\hat{p}_2,..., \hat{x}_n, \hat{p}_n)$, satisfying the canonical commutation relations \cite{arvind1995}
\begin{align*}
[\hat{Q}_j,\hat{Q}_k]=i\Omega_{jk}, ~~~\Omega=\mathbb{1}_n\otimes
\begin{pmatrix}
0 & 1 \\
-1 & 0
\end{pmatrix},
\end{align*}
where $\mathbb{1}_n$ is the $n \times n$ identity matrix. Transformations of coordinates that preserve the canonical commutation relation can be represented by symplectic transformation matrices $S$ such that $S\Omega S^\text{T}=\Omega$.

Gaussian states are a special class of continuous-variable states. They are defined as the states whose Wigner function is a Gaussian distribution \cite{ferraro2005, wang2007, weedbrook2012, adesso2014, serafini2017}.
It is known that an arbitrary Gaussian state $\hat{\rho}$ can be written in the Gibbs-exponential form as \cite{banchi2015, holevo2011}
\begin{align}\label{gibbs}
\hat{\rho}_{\text{Gibbs}}[G, u]\equiv\exp\left[-\frac{1}{2}(\hat{Q}-u)^{\text{T}} G(\hat{Q}-u)\right]/Z_V,
\end{align}
where $u=\text{Tr}[\hat{\rho}\hat{Q}]$ is the first moment vector, $G$ is the Gibbs matrix defined as $G=2i\Omega\coth^{-1}(2Vi\Omega)$ with the covariance matrix $V_{jk}=\text{Tr}[\hat{\rho}\{\hat{Q}_j-u_j,\hat{Q}_k-u_k\}_+]/2$, and $Z_V=\det(V+i\Omega/2)^{1/2}$ is a normalization factor which we omit throughout this work for convenience.
The Gibbs-exponential form of Eq.~\eqref{gibbs} makes it easy to deal with the square root of the density matrices, e.g., in Eq.~\eqref{meq}.\\

Let us substitute two arbitrary Gaussian states $\hat{\rho}_{j} (j=0,1)$, characterized by $u_j$ and $G_j$ through Eq.~\eqref{gibbs}, to the operator $\hat{M}$ of Eq.~\eqref{meq} in order to find the optimal measurement for quantum fidelity between Gaussian states. As the first main result of this work, we find, after some algebra (see Appendix A for the detail), that the operator $\hat{M}$ takes the exponential 
form, written up to an unimportant normalization factor as
\begin{align}
	\hat{M}\propto\hat{D}(u_1)\exp\left[-\frac{1}{2}\hat{Q}^{\text{T}} G_{\text{M}}\hat{Q}- v_\text{M}^{\text{T}}\hat Q\right]\hat{D}^\dagger(u_1),
	\label{Mgaussian}
\end{align}
where the matrix $G_{\text{M}}$ is the solution of the equation
\begin{equation}
	e^{i\Omega G_{\text{M}}} e^{i\Omega G_1} e^{i\Omega G_{\text{M}}} = e^{i\Omega G_0},
	\label{GMgen}
\end{equation}
$\hat{D}(u)=e^{-u^\text{T}i\Omega\hat{Q}}$ is the displacement operator, and $v_{M}$ is a real vector, which can be explicitly expressed for particular cases as below. 
Note that $G_{\text{M}}$ is not necessarily positive definite, unlike $G_0$ and $G_1$ characterizing Gaussian states, indicating that the operator $\hat{M}$ may not be written in the form of a Gaussian state depending on the feature of $G_\text{M}$.

When the Gibbs matrices of two multimode Gaussian states are equal, i.e., $G_0=G_1=(S^{-1})^\text{T}[\oplus_{j=1}^n g_j\mathbb{1}_2] S^{-1}$ with $g_{j}$ being the symplectic spectrum, Eq.~\eqref{GMgen} has a trivial solution $G_{\text{M}}=0$, allowing Eq.~\eqref{Mgaussian} to take a simpler form of $\hat{M}\propto e^{v_\text{M}^{\text{T}}(\hat Q-u_1)}$ where $v_\text{M}=(S^{-1})^{\text{T}}[\oplus_{j=1}^n \tanh(g_j/2)\mathbb{1}_2] (u_0-u_1)$. The eigenbasis of the operator $\hat{M}$ is thus that of a quadrature operator followed by a unitary operator $\hat{D}(u_1)$, which is overall still a quadrature operator.
When $G_0\neq G_1$, on the other hand, the operator $\hat{M}$ of Eq.~\eqref{Mgaussian} reduces to
\begin{align}
\hat{M}&\propto \hat{D}(u_1)\hat D(u_\text{M})\hat{\rho}_{\text{Gibbs}}[G_{\text{M}}, 0]\hat D^\dagger(u_\text{M})\hat{D}^\dagger(u_1)
\label{MgaussianD}
\end{align}
where $v_\text{M}=G_{\text{M}}u_\text{M}$ is used and the expression of $u_\text{M}$ is provided in Appendix A. Note that $v_\text{M} =0$ for equal displacements ($u_0=u_1$). 
When $G_0$ and $G_1$ are diagonalized by the same symplectic matrix $S$, individual modes of the states can be completely decoupled to be a product of single-mode states by applying a Gaussian unitary operation $\hat{U}_S$ corresponding to $S$. We thus investigate the single-mode case more intensively in the following section.

It is known that the Gibbs matrices are singular when symplectic eigenvalues of the covariance matrix are equal to $1/2$~\cite{banchi2015}. The continuity of the above expression enables the singular case to be treated as a limiting case. To this end, we replace the singular symplectic eigenvalues by $1/2+\epsilon$ with a small positive $\epsilon$, so that Eq.~\eqref{GMgen} is well defined as
\begin{equation}
	e^{i\Omega G_{\text{M}}}  = 
	e^{-i\Omega G_1/2} 
	\sqrt{ e^{i\Omega G_1/2} e^{i\Omega G_0} e^{i\Omega G_1/2} }
	e^{-i\Omega G_1/2}.
	\label{GMsol}
\end{equation}
In the limit $\epsilon\to0$, the above expression leads to an optimal measurement, but note that the optimal measurement may not be unique when rank-deficient states are involved.

\section{Optimal measurements for single-mode Gaussian states}
The operator $\hat{M}$ of Eq.~\eqref{Mgaussian}, whose eigenstates constitute the POVM elements of the optimal measurement, can be analyzed for specific cases of interest. Here we concentrate on single mode Gaussian states, exhibiting rich physics and the immediate relevance to quantum parameter estimation as will be discussed in the next section.
An arbitrary single-mode Gaussian state can be written as 
\begin{align*}
\hat{\rho}&=\hat{D}(u)\hat{S}(\xi)\hat{\rho}_{\text{T}}\hat{S}^\dagger(\xi)\hat{D}^\dagger(u),
\end{align*}
where $\hat{\rho}_\text{T}=\sum_{n=0}^\infty\bar{n}^n/(\bar{n}+1)^{n+1}|n\rangle\langle n|$ is a thermal state with the average number of thermal quanta $\bar{n}$, and $\hat{S}(\xi)$ is a squeezing operator with a squeezing parameter $\xi\equiv re^{i\theta_s} \in \mathbb{C}$. Note that when $\theta_s=0$, the Gibbs matrix in Eq.~\eqref{gibbs} is written as 
\begin{align}\label{GibbsSM}
G=2\coth^{-1}(2\bar{n}+1)
\begin{pmatrix}
e^{2r} & 0 \\
0 & e^{-2r}
\end{pmatrix}.
\end{align}

For two given arbitrary single-mode Gaussian states, one can always find a Gaussian unitary operator $\hat{V}$ which transforms one state to a thermal state and, accordingly, the other state to a general Gaussian state but squeezed in $\hat{x}$ or $\hat{p}$ and displaced by $u_0$. The property in Eq.~\eqref{Mproperty} thus makes it sufficient to consider, without loss of generality, two Gaussian states: a general state written as $\hat{\rho}_{0}=\hat{\rho}_\text{Gibbs}[G_0,u_0]$, with $G_0$ being a diagonal matrix as Eq.~\eqref{GibbsSM}, and a thermal state written as $\hat{\rho}_{1}=\hat{\rho}_\text{Gibbs}[G_1,u_1=0]$, where $G_1=g_1 \mathbb{1}_{2}$ with $g_1\equiv 2\coth^{-1}(2\bar{n}_1+1)$. 


Let us first consider the case that $\hat{\rho}_0$ and $\hat{\rho}_1$ are full-rank states, 
i.e., 
$\bar{n}_j\neq0$ for both $j=0,1$.
For the states with $G_0=G_1$, one can easily show that $\hat{M}\propto e^{v_\text{M}^{\text{T}}\hat Q}$, where $v_\text{M}=\tanh(g_1/2)(S^{-1})^{\text{T}} u_0$, and its eigenbasis is that of a quadrature operator as in the multimode case. When ${G}_0\neq {G}_1$, on the other hand, the operator of $\hat{M}$ can be expressed as $\hat{M} \propto \hat V \hat{D}(u_\text{M})\hat{\rho}_{\text{Gibbs}}[G_{\text{M}},0]\hat{D}^\dagger(u_\text{M}) \hat V^\dagger$, similar to Eq.~\eqref{MgaussianD}.
The identification of optimal measurements requires the operator of $\hat{M}$ to be diagonalized, which boils down to a diagonalization of $\hat{\rho}_{\text{Gibbs}}[G_{\text{M}},0]$ for which the feature of the matrix $G_{\text{M}}$, not necessarily positive definite, matters.
Interestingly, it turns out that the type of the optimal measurements or that of the eigenbasis of the operator $\hat{\rho}_{\text{Gibbs}}[G_{\text{M}},0]$ can be simply classified by the signs of eigenvalues, $d_{1}$ and $d_{2}$, of the matrix $G_{\text{M}}$. The identified types are listed below as the second main result of this work.

\begin{figure}[t]
\centering
\includegraphics[width=0.4\textwidth]{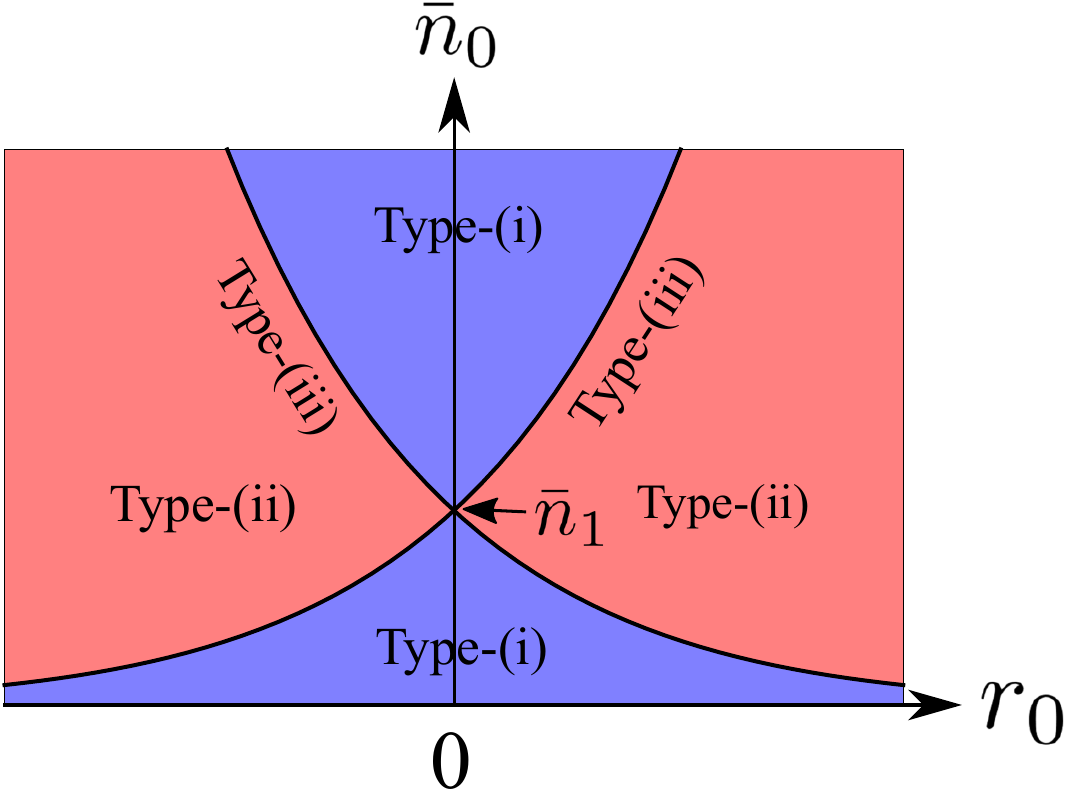}
\caption{Classification of optimal measurements as a function of $r_{0}$ and $\bar{n}_{0}$ for a given $\bar{n}_{1}$. The regions where \text{type (i)} and \text{type (ii)} are optimal are divided by the the black curves, at which  \text{type (iii)} is optimal, also including the intersection point when the Gibbs matrices of the states are identical.}
\label{plot1}
\end{figure}

\begin{itemize}
\item [(i)] If the signs of the eigenvalues of $G_{\text{M}}$ are the same (i.e., $d_{1}d_{2}>0$), i.e., $G_{\text{M}}$ is positive definite or negative definite, then the eigenbasis of $\hat{M}$ is that of the number operator $\hat{n}=(\hat{x}^2+\hat{p}^2-1)/2$ followed by the unitary operation $\hat{V}$ and a squeezing operation that makes the magnitude of the eigenvalues the same. Thus, an excitation-number-resolving detection is the optimal measurement.
\item [(ii)] If the signs of the eigenvalues are different (i.e., $d_{1}d_{2}<0$), then the eigenbasis of $\hat{M}$ is that of $\hat{x}\hat{p}+\hat{p}\hat{x}$ followed by a similar unitary operation to the one considered in type (i). Hence, a measurement scheme performing projection onto the eigenbasis of $\hat{x}\hat{p}+\hat{p}\hat{x}$ is the optimal measurement.
\item [(iii)] If only one of the eigenvalues is zero (i.e., either $d_{1}=0$ or $d_{2}=0$), then the eigenbasis of $\hat{M}$ is that of a quadrature operator along a certain direction. Therefore, homodyne detection is the optimal measurement.
\end{itemize}
Note that the optimal measurement of type (ii) is a definitely non-Gaussian measurement~\cite{oh2018}, the implementation of which is unfortunately unknown. The eigenvalues can be found by solving Eq.~\eqref{GMgen} and written as a function of the squeezing parameter $r_0$, and thermal quanta $\bar{n}_{0}$ and $\bar{n}_{1}$ (see Appendix B for the detail). It enables mapping the above classification to the parameter space of $r_0$, $\bar{n}_{0}$ and $\bar{n}_{1}$, as depicted in Fig.~\ref{plot1} for a given $\bar{n}_{1}$. 
The case that $G_0=G_1$, where type (iii) is optimal, is also represented by the intersection point where $\bar{n}_0=\bar{n}_1$.
Thus, the diagram shown in Fig.~\ref{plot1} covers all pairs of single-mode Gaussian states through the Gaussian unitary operator $\hat{V}$.

It is worth discussing special cases, when each type is optimal. First, consider the case when $\hat{\rho}_0$ is a displaced thermal state. Thus, $G_0=\text{diag}(g_0,g_0)$, which corresponds to the case where the distinct Gibbs matrices of two Gaussian states are diagonalized by the same symplectic transformation. In this case, Eq.~\eqref{GMgen} leads to $\hat{\rho}_{\text{Gibbs}}[G_{\text{M}},0]=\exp\big[-\frac{1}{2}(g_1-g_0)\hat{Q}^{\text{T}} \hat{Q}\big]$, and the eigenbasis of $\hat{M}$ is the number basis followed by $\hat{V}$ and $\hat{D}(u_\text{M})$. Hence, \text{type (i)} is optimal. This result can also be inferred by the fact that the same unitary operation diagonalizes both states into thermal states, and their eigenbasis is the number state. Second, consider the case when $\bar{n}_0=\bar{n}_1$ and $G_0$ has distinct eigenvalues, i.e., $\hat{\rho}_0$ is a squeezed state. It renders the signs of $d_1$ and $d_2$ different regardless of $r_{0}$ and $\bar{n}_{0}=\bar{n}_{1}$, i.e., \text{type (ii)} is optimal. Third, consider the case that the amount of nonzero squeezing $r_0$ obeys a certain ratio of functions of thermal quanta $\bar{n}_0$ and $\bar{n}_1$, given as
\begin{align}\label{homcon1}
e^{\pm 2r_0}=\frac{\bar{n}_0(\bar{n}_0+1)(2\bar{n}_1+1)}{\bar{n}_1(\bar{n}_1+1)(2\bar{n}_0+1)},
\end{align}
where the signs $\pm$ in the exponent correspond to the cases of $d_{2}=0$ and $d_{1}=0$, respectively. 
When $d_{2}=0$, the operator $\hat{M}$ is simply written as $\hat{M}\propto\hat V\hat{D}(u_\text{M})\exp\big[-\frac{d_1}{2}\hat{x}^2\big]\hat{D}^\dagger(u_\text{M})\hat V^\dagger$ and thus \text{type (iii)} with the quadrature measurement of $\hat{x}$ is optimal, whereas \text{type (iii)} with the quadrature measurement of $\hat{p}$ is optimal when $d_{1}=0$.


Now consider the case of rank-deficient Gaussian states. Since all rank-deficient Gaussian states are a pure state and the inverse of a pure state does not exist, $\hat{M}$ of Eq.~\eqref{meq} needs to be treated with care. Assuming $\hat{\rho}_1$ to be a pure state without loss of generality and projecting both $\hat{\rho}_{0}$ and $\hat{\rho}_{1}$ onto the support of $\hat{\rho}_1$, where their inverses can be defined, one can write the operator $\hat{M}$ of Eq.~\eqref{meq} as
\begin{align*}
\hat{M}=\hat{\rho}_1^{-1/2}\sqrt{\hat{\rho}_1^{1/2}\hat{\Pi}_1\hat{\rho}_0\hat{\Pi}_1\hat{\rho}_1^{1/2}}\hat{\rho}_1^{-1/2},
\end{align*}
where $\hat{\Pi}_1$ is the projector onto the support of $\hat{\rho}_1$~\cite{wilde2017}.
For $\hat{\rho}_1=|\psi_1\rangle\langle\psi_1|$ and, consequently, $\hat{\Pi}_1=|\psi_1\rangle\langle \psi_1|$, it is therefore clear that $\hat{M}\propto |\psi_1\rangle\langle \psi_1|$.
The same result can also be derived by considering pure states as a limiting case of zero temperature (see Appendix C for the detail). Thus, the optimal POVM set is $\{|\psi_1\rangle\langle\psi_1|,\mathbb{1}-|\psi_1\rangle\langle\psi_1|\}$ and can be implemented by applying the Gaussian unitary transformation $\hat{S}^\dagger(\xi_1) \hat{D}^\dagger(u_1)$ that transforms $\hat{\rho}_1$ to a vacuum state, followed by performing on-off detection.
It is worth emphasizing again that the optimal measurement offered by the operator $\hat{M}$ for pure states is not unique, so that the suggested setup is merely one of the optimal measurements, all satisfying the conditions of Eqs.~\eqref{fidc1} and \eqref{fidc2}.

\section{Relevance to quantum metrology}
Quantum parameter estimation is an informational task to estimate an unknown parameter $\theta$ of interest by using quantum systems~\cite{helstrom1976, paris2009}. In the standard scenario of quantum parameter estimation, $N$ independent copies of quantum states that contain information about an unknown parameter are measured by a POVM and the estimation is performed by manipulating the measurement data. The ultimate precision bound of the estimation is governed by quantum Cram\'{e}r-Rao inequality, stating that the mean square error of any unbiased estimator is lower bounded by the inverse of quantum Fisher information multiplied by the number of copies $N$~\cite{helstrom1976}. Thus, quantum Fisher information is the most crucial quantity which determines the ultimate precision of estimation~\cite{braunstein1994}, which is written as
\begin{align*}
H(\theta)=\text{Tr}[\hat{\rho}_\theta\hat{L}_\theta^2],
\end{align*}
where $\hat{L}_\theta$ is the symmetric logarithmic derivative (SLD) operator satisfying $\partial\hat{\rho}_\theta/\partial\theta=\hat{\rho}_\theta\hat{L}_\theta+\hat{L}_\theta\hat{\rho}_\theta$.

The quantum Fisher information $H(\theta)$ can be written in terms of quantum fidelity $F(\hat{\rho}_\theta,\hat{\rho}_{\theta+d\theta})$ as~\cite{banchi2015}
\begin{align*}
H(\theta)=\frac{4[1-F(\hat{\rho}_\theta,\hat{\rho}_{\theta+d\theta})]}{d\theta^2}.
\end{align*}
This implies that quantum parameter estimation is related to distinguishing two infinitesimally close states $\hat{\rho}_\theta$ and $\hat{\rho}_{\theta+d\theta}$.
Indeed, similar to quantum fidelity, quantum Fisher information is defined as the maximal classical Fisher information over all possible POVMs, and the optimal POVM $\{\hat{E}_x\}$ has to satisfy~\cite{braunstein1994}
\begin{align}
\hat{E}_x^{1/2}(\hat{\rho}_\theta^{1/2}-\lambda_x\hat{L}_\theta\hat{\rho}_\theta^{1/2})=0, \label{qfic1}\\
\text{Tr}[\hat{E}_x \hat{\rho}_\theta\hat{L}_\theta]\in \mathbb{R}, \label{qfic2}
\end{align}
where $\lambda_x$ is a constant.
It is known that the projection onto the eigenbasis of $\hat{L}_\theta$ is the optimal measurement for quantum Fisher information~\cite{braunstein1998}. This means that the SLD operator plays the same role as the operator $\hat{M}$ does for quantum fidelity. We prove that the above conditions of Eqs.~\eqref{qfic1} and \eqref{qfic2} are indeed equivalent to the conditions of Eqs.~\eqref{fidc1} and \eqref{fidc2}, resulting in the relation for arbitrary quantum states $\hat{\rho}_\theta$ and $\hat{\rho}_{\theta+d\theta}$,
\begin{align}
\hat{M}(\hat{\rho}_\theta,\hat{\rho}_{\theta+d\theta})\simeq 1 + \hat{L}_\theta d\theta/2
\end{align}
for infinitesimal $d\theta$ (see Appendix D for the proof). This indicates that the optimal POVM for quantum fidelity between $\hat{\rho}_\theta$ and $\hat{\rho}_{\theta+d\theta}$ offers the optimal measurement for quantum parameter estimation, yielding the maximal Fisher information, i.e., quantum Fisher information. 

Especially for Gaussian states, since the matrix $G_{\text{M}}$ and the vector $v_\text{M}$ are infinitesimal for $\hat{\rho}_\theta$ and $\hat{\rho}_{\theta+d\theta}$, and thus
\begin{align*}
\hat{M}(\hat{\rho}_\theta,\hat{\rho}_{\theta+d\theta})\simeq 1-\hat{D}(u_\theta)(\hat{Q}^\text{T}G_{\text{M}}\hat{Q}/2-v_\text{M}^\text{T}\hat{Q})\hat{D}^\dagger(u_\theta),
\end{align*}
the SLD operator is simply written as 
\begin{align}\label{SLD}
\hat{L}_\theta d\theta=-\hat{D}(u_\theta)(\hat{Q}^\text{T}G_{\text{M}}\hat{Q}-2v_\text{M}^\text{T}\hat{Q})\hat{D}^\dagger(u_\theta)+\nu,
\end{align}
where $\nu=\tr{\hat{D}^\dagger(u_\theta)\hat{\rho}_\theta\hat{D}(u_\theta){\hat{Q}^\text{T}G_{\text{M}}\hat{Q}}}$ can be determined from $\tr{\hat{\rho}\hat{L}_\theta}=0$.
Taking an infinitesimal limit in Eq.~\eqref{GMgen}, one can show that $G_{\text{M}}$ for an infinitesimal $d\theta$ is the solution of
\begin{align}
4V_\theta G_{\text{M}} V_\theta+\Omega G_{\text{M}} \Omega+2d\theta\frac{\partial V_\theta}{\partial\theta}=0,
\end{align}
and is formally written in a basis-independent form as
\begin{align}\label{GMsol2}
G_{\text{M}}=i\Omega\sum_{m=0}^{\infty} W_\theta^{-m-1}\frac{\partial W_\theta}{\partial \theta}W_\theta^{-m-1}d\theta.
\end{align}
and $v_\text{M}=V_\theta^{-1}({\partial u_\theta}/\partial{\theta}) d\theta/2$.
Here, $u_\theta$ and $V_\theta$ are the first moment vector and the covariance matrix of $\hat{\rho}_\theta$, respectively, and $W_\theta=-2V_\theta i\Omega $. The derivation of $G_{\text{M}}$ and $v_\text{M}$ is provided in Appendix E. The relation of $\hat{M}$ and $\hat{L}_\theta$, and the expressions of $G_{\text{M}}$ and $v_\text{M}$, enable one to find the SLD operator $\hat{L}_\theta$ directly from the operator $\hat{M}$.
Finally, from the SLD operator, one can easily derive the expression of the quantum Fisher information:
\begin{align}
H(\theta)=-\text{Tr}\bigg[\frac{\partial V_\theta}{\partial \theta} G_{\text{M}}\bigg]+\frac{\partial u_\theta}{\partial{\theta}} V^{-1}_\theta \frac{\partial u_\theta}{\partial{\theta}}.
\end{align}
The derivation is provided in Appendix D.
As a remark, note that the expressions of $G_{\text{M}}$, $v_\text{M}$ and quantum Fisher information are equivalent to those found in Refs.~\onlinecite{serafini2017, jiang2014}, but our derivation based on quantum fidelity is significantly simpler and more straightforward. Furthermore, replacing a single parameter $\theta$ by a vector of multiparameter $\vec{\theta}$ and defining the SLD operators $\hat{L}_{\theta_j}$ by $\partial\hat{\rho}_{\vec{\theta}}/\partial\vec{\theta}_j=\hat{\rho}_{\vec{\theta}}\hat{L}_{\theta_j}+\hat{L}_{\theta_j}\hat{\rho}_{\vec{\theta}}$, the expression of the quantum Fisher information matrix $H_{jk}(\vec{\theta})=\tr{\hat{\rho}_{\vec{\theta}}\{\hat{L}_{\theta_j}, \hat{L}_{\theta_k}\}_+}$ can be easily derived by using a similar method \cite{nichols2018, safranek2019}.

In the following sections, we find optimal measurements for displacement, phase, squeezing, and loss parameter estimation in relation to our results for quantum fidelity.

\subsection{Displacement parameter estimation}
For a single-mode Gaussian probe state $\hat{\rho}$, the displacement operation $\hat{D}(\alpha)$ only changes the first moment while keeping the second moments fixed:
\begin{align*}
u\rightarrow u+(\alpha, 0)^\text{T}, ~~~V\rightarrow V,
\end{align*}
where $\alpha\in\mathbb{R}$ is assumed without loss of generality. Therefore, the first moment vectors and the covariance matrices of $\hat{\rho}_\alpha$ and $\hat{\rho}_{\alpha+d\alpha}$ are related as
\begin{align*}
u_{\alpha+d\alpha}=u_{\alpha}+(d\alpha,0)^\text{T},~~~V_{\alpha+d\alpha}=V_{\alpha},
\end{align*}
respectively. Since the covariance matrix is invariant, corresponding to the case of the intersection point in Fig.~\ref{plot1}, one can immediately see that the optimal measurement for quantum fidelity between $\hat{\rho}_\alpha$ and $\hat{\rho}_{\alpha+d\alpha}$ is \text{type (iii)}, so that the optimal measurement for estimation of the displacement parameter $\alpha$ is also \text{type (iii)}.
Explicitly, using the expression of $v_\text{M}$, one can easily obtain the SLD operator and quantum Fisher information,
\begin{align*}
\hat{L}_\alpha&=\hat{D}(u_\alpha)([V^{-1}_\alpha]_{11}\hat{x}+[V^{-1}_\alpha]_{12}\hat{p})\hat{D}^\dagger(u_\alpha) \\
&=[V^{-1}_\alpha]_{11}(\hat{x}-u_\alpha)+[V^{-1}_\alpha]_{12}\hat{p}, \\
H(\alpha)&=[V^{-1}_\alpha]_{11}.
\end{align*}
Thus, the optimal measurement is homodyne detection as expected.

\subsection{Phase parameter estimation}
Let us consider a single-mode Gaussian probe state $\hat{\rho}$ that undergoes a phase shifter $\hat{R}(\theta)=e^{-i\theta\hat{Q}^\text{T}\hat{Q}/2}$ with a phase parameter $\theta$ to be estimated. Since the displacement operation performed to the probe state can be factored out as shown in Eq.~\eqref{MgaussianD}, we focus on only the state with zero mean for simplicity, i.e.,
\begin{align*}
\hat{\rho} \rightarrow \hat{\rho}_\theta=\hat{R}(\theta)\hat{S}(\xi)\hat{\rho}_\text{T}\hat{S}^\dagger(\xi)R^\dagger(\theta).
\end{align*}
The relevant states under investigation are $\hat{\rho}_{\theta}$ and $\hat{\rho}_{\theta+d\theta}$, but the full expressions with an arbitrary angle $\theta$ get involved without altering the type of optimal measurement. We thus focus on the states $\hat{\rho}_{\theta}$ and $\hat{\rho}_{\theta+d\theta}$ at $\theta=0$ and further assume $\hat{\rho}_0$ and $\hat{\rho}_{d\theta}$ to be the $p$-squeezed thermal state and a rotated squeezed thermal state, respectively, without loss of generality. 

Let us proceed with $\hat{\rho}_{0}$ and $\hat{\rho}_{\theta}$ first, and then take the limit $\theta\rightarrow0$ at the end. The covariance matrices of $\hat{\rho}_{0}$ and $\hat{\rho}_{\theta}$ are, respectively, written as
\begin{align*}
V_0&\propto
\begin{pmatrix}
e^{2r} & 0 \\
0 & e^{-2r}
\end{pmatrix}, \\
V_\theta&\propto
\begin{pmatrix}
\cosh2r+\cos2\theta\sinh2r & \sinh2r\sin2\theta \\
\sinh2r\sin2\theta & \cosh2r-\cos2\theta\sinh2r
\end{pmatrix},
\end{align*}
where the proportionality becomes an equality with adding a prefactor of $(2\bar{n}+1)/2$. Through the Gaussian unitary operation $\hat{V}$, these states are transformed to a squeezed thermal state and a thermal state with the same number of thermal quanta. Thus, one may immediately infer from Fig.~\ref{plot1} that the optimal measurement is \text{type (ii)} regardless of $\theta$. Let us see if this is indeed the case. For the states $\hat{\rho}_{0}$ and $\hat{\rho}_{\theta}$, it can be shown that  
\begin{align*}
G_{\text{M}}= A
\begin{pmatrix}
-\sin\theta & \cos\theta\\
\cos\theta & \sin\theta
\end{pmatrix},
\end{align*}
where a constant $A$ is given such that $\cos A =(4\bar{n}^2+4\bar{n}+2)/[(4\bar{n}^2+2\bar{n}+1)(4\bar{n}^2+6\bar{n}+3)+(2\bar{n}+1)^2\cos2\theta+2(2\bar{n}+1)^2\cosh4r\sin^2\theta]^{1/2}$.
Since the eigenvalues of $G_\text{M}$ are different, the optimal measurement for quantum fidelity between $\hat{\rho}_{0}$ and $\hat{\rho}_{\theta}$ is \text{type (ii)}.
To apply this to quantum Fisher information, we take the limit $\theta\rightarrow 0$, resulting in 
\begin{align*}
G_{\text{M}}=\frac{(2\bar{n}+1)\sinh{2r}}{2\bar{n}^2+2\bar{n}+1}d\theta
\begin{pmatrix}
0 & 1 \\
1 & 0
\end{pmatrix}.
\end{align*}
Hence,
\begin{align}\label{phaseM}
\hat{M}=1-\frac{(2\bar{n}+1)\sinh{2r}}{2(2\bar{n}^2+2\bar{n}+1)}d\theta(\hat{x}\hat{p}+\hat{p}\hat{x})=1+\hat{L}_\theta d\theta/2,
\end{align}
where $\hat{L}_\theta$ is the SLD operator in phase estimation \cite{oh2018}. This reveals that the operators $\hat{M}$ and $\hat{L}_\theta$ have the common eigenbasis. It is now clear that the optimal measurement for phase parameter estimation is \text{type (ii)}, as also recently found via the SLD operator in Ref.~\onlinecite{oh2018}.
Also note that while the above result is derived by an explicit optimal measurement for quantum fidelity, the same result can be easily derived by using Eq.~\eqref{GMsol2}.

\subsection{Squeezing parameter estimation}
We consider squeezing parameter estimation with an arbitrary Gaussian state as a probe state,
\begin{align*}
\hat{\rho} \rightarrow \hat{\rho}_\zeta=\hat{S}(\zeta)\hat{D}(u)\hat{S}(\xi)\hat{\rho}_\text{T}\hat{S}^\dagger(\xi)\hat{D}^\dagger(u)S^\dagger(\zeta),
\end{align*}
where we assume $\zeta\in\mathbb{R}$ for simplicity. It corresponds to the case when we estimate the strength of the squeezing parameter along the $\hat{p}$~axis.
Since $\hat{\rho}_\zeta$ and $\hat{\rho}_{\zeta+d\zeta}$ have different squeezing parameters under the same average number of thermal quanta, just like the case of phase estimation, the optimal measurement is \text{type (ii)}.
Indeed, one can derive the SLD operator using Eq.~\eqref{GMsol2},
\begin{align*}
\hat{L}_\theta&=\frac{2\bar{n}+1}{2\bar{n}^2+2\bar{n}+1}\hat{D}(u)\hat{Q}^\text{T}\\
&\quad\times\text{diag}\big[e^{-2\zeta}(\cosh{2r}-\cos{\theta_s}\sinh{2r}),\\
&\quad\quad\quad\quad\quad -e^{2\zeta}(\cosh{2r}+\cos{\theta_s}\sinh{2r})\big]\hat{Q}\hat{D}^\dagger(u)+\nu,
\end{align*}
which is clearly type (ii) because the signs of eigenvalues of $G_{\text{M}}$ are different.
Quantum Fisher information can also be easily obtained as
\begin{align*}
H(s)&=\frac{2(2 \bar{n}+1)^2}{2 \bar{n}^2+2\bar{n}+1}(\cosh^2 2r-\cos^2 \theta_s \sinh^2 2r) \\
&~~\quad\quad+\frac{4 |\alpha| ^2 }{2 \bar{n}+1}(\cosh 2 r-\sinh 2 r \cos (2 \theta_c+\theta_s)),
\end{align*}
where we have defined $u=\sqrt{2}|\alpha|(\cos\theta_c, \sin\theta_c)^\text{T}$ and $\theta_c$ is the displacement angle.

\subsection{Loss parameter estimation}
Consider a single-mode Gaussian probe state $\hat{\rho}$ that undergoes a phase-insensitive loss channel, and the dynamics of the state is described by the quantum master equation as
\begin{align} \label{loss}
\frac{d\hat{\rho}}{dt}=\frac{\gamma}{2}(2\hat{a}\hat{\rho} \hat{a}^\dagger-\hat{a}^\dagger \hat{a}\hat{\rho}-\hat{\rho} \hat{a}^\dagger \hat{a}),
\end{align}
where $\hat{a}=(\hat{x}+i\hat{p})/\sqrt{2}$ is the annihilation operator and $\gamma$ is the loss rate to be estimated. The solution of the above differential equation for a single-mode Gaussian probe state can be given in terms of the first moment vector and the covariance matrix as \cite{ferraro2005}
\begin{align*}
u_{t=0}& \rightarrow u_t=e^{-\gamma t/2}u_0,\\ 
V_{t=0}&\rightarrow V_t=e^{-\gamma t}V_0+(1-e^{-\gamma t})\mathbb{1}_2/2.
\end{align*}
Note that the dynamics of the covariance matrix does not change the symplectic transformation diagonalizing the covariance matrix. Therefore, the Gaussian unitary operation $\hat{V}$ may transform these states to two thermal states with different number of thermal quanta. It is thus clear from Fig.~\ref{plot1} that the optimal parameter for quantum fidelity between $\hat{\rho}_{\gamma}$ and $\hat{\rho}_{\gamma+d\gamma}$ is \text{type (i)}, so the optimal measurement for the loss parameter estimation is also \text{type (i)}. 
Specifically, one can easily obtain that
\begin{align*}
G_{\text{M}}&=A\times\text{diag}(\sin^4\phi -e^{-2 r} \cos ^4\phi, \\
&\quad\quad\quad\quad\quad\quad\quad\sin ^4\phi-e^{2 r} \cos ^4\phi )t d\gamma, \\
H(\gamma)&=\frac{\cos^2\phi(1-2 \sin^2\phi \cos^2\phi)\sinh^2{r}}{\sin^2\phi(1+2\sin^2\phi\cos^2\phi\sinh^2{r})}t^2,
\end{align*}
where we have defined $\cos^2\phi=e^{-\gamma t}$ and $A=4/(-2 \sinh ^2r \cos 4 \phi +\cosh 2 r+7)\sin^2\phi$ and zero-mean input states are assumed for simplicity. 
The matrix $G_{\text{M}}$ is obviously negative definite; thus it corresponds to \text{type (i)}. This reproduces the result in Refs.~\onlinecite{monras2007, pinel2013}. The optimality of \text{type (i)} holds also for other phase-insensitive loss parameter estimations as long as the symplectic matrix that diagonalizes the covariance matrix remains the same with loss parameter $\gamma$ or time $t$.

\section{Discussion}
We have found the optimal POVMs for quantum fidelity between two multi-mode Gaussian states in a closed analytical form. The full generality of our result has allowed us to further elaborate on the case of single-mode Gaussian states in depth. We have demonstrated that there exist only three different types of optimal measurements, along with Gaussian unitary operations.
An excitation-number-counting measurement is optimal when the covariance matrices of the states are diagonalized by the same symplectic matrix, while the projection onto the eigenbasis of $\hat{x}\hat{p}+\hat{p}\hat{x}$ is optimal when the average numbers of thermal quanta of two quantum states are equal. While there exist other cases where the aforementioned optimal measurements are, respectively, optimal, the optimality of the quadrature measurement holds only for two cases: when the covariance matrices are the same or when the squeezing strength of $\hat{\rho}_0$ is equal to a particular ratio, represented in Eqs.~\eqref{homcon1}, of thermal quanta contributions between the two states.

We have also shown the relevance of the optimal measurement for quantum fidelity to quantum parameter estimation. We have proven the equivalence between the optimal measurement for quantum fidelity and that for quantum Fisher information, enabling one to readily derive optimal measurements for quantum parameter estimation using Gaussian states. We expect our approach, based on the fundamental relation we proved, to pave a way to study quantum parameter estimation or other quantum information processing.

A particularly interesting potential application of our optimal measurements is quantum hypothesis testing \cite{chefles2000, helstrom1976, barnett2009, pirandola2018}.
The minimal error probability of quantum state discrimination is given by the Helstrom bound, achieved only by the Helstrom measurement \cite{helstrom1976}.
However, finding a closed form of the Helstrom measurement for Gaussian states is generally challenging.
The quantum fidelity is known to set an upper bound for the error of quantum state discrimination \cite{fuchs1999, barnum2002, montanaro2008}, and the optimal measurement for quantum fidelity enables one to lower the error of particular schemes such as the maximum-likelihood test \cite{cover2012}.
In this context, one could address the question of whether the optimal measurements we have found can be exploited for variants of quantum state discrimination such as quantum illumination \cite{lloyd2008, tan2008} and quantum reading \cite{pirandola2011}.

While the excitation-number-resolving detection and the quadrature variable measurement are experimentally feasible with current technology, the measurement setup projecting onto the eigenbasis of the operator $\hat{x}\hat{p}+\hat{p}\hat{x}$ is not yet known. We hope that an appropriate measurement setup will be constructed in the near future in response to the significance having arisen not only from this work but also from the recent study for phase estimation~\cite{oh2018}. We also leave further classification of the optimal measurements for multi-mode Gaussian states as future work, which can be straightforwardly made from our results at the expense of increased complexity. 

\begin{acknowledgements}
C.O. and H.J. are supported by a National Research Foundation of Korea grant funded by the Ministry of Science and ICT (Grant No. 2010-0018295 and 2018K2A9A1A06069933).
L.B. was supported by the UK EPSRC Grant No. EP/K034480/1.
S.-Y.L. is supported by Basic Science Research Program through the National Research Foundation of Korea (NRF) funded by the Ministry of Education (Grant No. 2018R1D1A1B07048633).
\end{acknowledgements}

\section*{Appendix}
\subsection{Simplification of the operator $\hat{M}$}
\setcounter{equation}{0}
\renewcommand{\theequation}{A\arabic{equation}}
Here, we simplify the operator $\hat{M}=\hat{\rho}_1^{-1/2}\sqrt{\hat{\rho}_1^{1/2}\hat{\rho}_0\hat{\rho}_1^{1/2}}\hat{\rho}_1^{-1/2}$ with $\hat{\rho}_0=e^{-\frac{(\hat{Q}-v_0)^{\text{T}}  G_0(\hat{Q}-v_0)}{2}}$ and $\hat{\rho}_1=e^{-\frac{\hat{Q}^\text{T} G_1 \hat{Q}}{2}}$.
Note that $e^{l^{\text{T}} i\Omega\hat{Q}}e^{-\hat{Q}^{\text{T}} G\hat{Q}/2}\propto e^{-(\hat{Q}-u)^{\text{T}} G(\hat{Q}-u)/2}$ with $u=(e^{-i\Omega G}-1)^{-1}l$, which is frequently used in this section.
Simplifying $\hat{\rho}_0$ in the following way,
\begin{align*}
\hat{\rho}_0=e^{\frac{1}{2}v_0^{\text{T}} i\Omega e^{-i\Omega G_0}v_0}e^{(e^{-i\Omega G_0}v_0-v_0)^{\text{T}}  i\Omega \hat{Q}} e^{-\frac{\hat{Q}^{\text{T}} G_0\hat{Q}}{2}} \propto e^{l_0^{\text{T}} i\Omega \hat{Q}}e^{-\frac{\hat{Q}^{\text{T}} G_0\hat{Q}}{2}}
\end{align*}
with $l_0= (e^{-i\Omega G_0}-1)v_0$, one can have
\begin{align*}
\hat{K}
=\hat{\rho}_1^{1/2}\hat{\rho}_0\hat{\rho}_1^{1/2}
\propto e^{-\frac{\hat{Q}^{\text{T}} G_1\hat{Q}}{4}}e^{l_0^{\text{T}} i\Omega \hat{Q}}e^{-\frac{\hat{Q}^{\text{T}} G_0\hat{Q}}{2}}e^{-\frac{\hat{Q}^{\text{T}} G_1\hat{Q}}{4}}.
\end{align*}
Bringing all the displacement operators to the left side, one can further simplify the matrix $\hat{K}$ as
\begin{align*}
\hat{K}
&\propto e^{k^{\text{T}} i\Omega \hat{Q}}e^{-\frac{\hat{Q}^{\text{T}} G_K\hat{Q}}{2}},
\end{align*}
where we have defined $k= e^{-i \Omega G_1/2}l_0$ and 
\begin{align}\label{Gk}
e^{-\frac{\hat{Q}^{\text{T}} G_K\hat{Q}}{2}}&= e^{-\frac{\hat{Q}^{\text{T}} G_1\hat{Q}}{4}}e^{-\frac{\hat{Q}^{\text{T}} G_0\hat{Q}}{2}}e^{-\frac{\hat{Q}^{\text{T}} G_1\hat{Q}}{4}}.
\end{align}
Defining $u_K$ as $(e^{-i\Omega G_K}-1)u_K= k$,
the operator $\hat{K}$ takes the Gibbs-exponential form, written as
\begin{align*}
\hat{K}\propto e^{-(\hat{Q}-u_K)^{\text{T}} G_K(\hat{Q}-u_K)/2},
\end{align*}
where $u_K$ is a real vector. The operator $\hat{M}=\hat{\rho}_1^{-1/2}\sqrt{\hat{K}}\hat{\rho}_1^{-1/2}$ can thus be written as
\begin{align*}
\hat{M}
&\propto e^{\frac{\hat{Q}^{\text{T}} G_1\hat{Q}}{4}}e^{l_1^{\text{T}} i\Omega \hat{Q}}e^{-\frac{\hat{Q}^{\text{T}} G_K\hat{Q}}{4}}e^{\frac{\hat{Q}^{\text{T}} G_1\hat{Q}}{4}},
\end{align*}
where $l_1=(e^{-i\Omega G_K/2}-1)u_K$. Again, we bring all the displacement operators to the left side,
\begin{align*}
\hat{M}
&\propto  e^{m^{\text{T}} i\Omega \hat{Q}}e^{-\frac{\hat{Q}^{\text{T}} G_{\text{M}}\hat{Q}}{2}},
\end{align*}
where $m= e^{i \Omega G_1/2}l_1$ and
\begin{align}\label{Gm}
e^{-\frac{\hat{Q}^{\text{T}} G_{\text{M}}\hat{Q}}{2}}=e^{\frac{\hat{Q}^{\text{T}} G_1\hat{Q}}{4}}e^{-\frac{\hat{Q}^{\text{T}} G_K\hat{Q}}{4}}e^{\frac{\hat{Q}^{\text{T}} G_1\hat{Q}}{4}}.
\end{align}

When $G_{\text{M}}=0$, corresponding to the case that $G_0=G_1$, we obtain $\hat{M}\propto e^{m^{\text{T}} i\Omega \hat{Q}}$, where $m=e^{i \Omega G_1/2}l_1$ is a pure imaginary vector. Especially if $G_0=G_1=\oplus_{j=1}^n g_j \mathbb{1}_2$, we obtain $m=-i[\oplus_{j=1}^n \tanh(g_j/2)\mathbb{1}_2]\Omega v_0$.
If $G_0=G_1$ are not diagonal, we introduce a symplectic transformation that diagonalizes the Gibbs matrices, $G_0=G_1= (S^{-1})^{\text{T}}[\oplus_{j=1}^n g_j\mathbb{1}_2]S^{-1}$, or, equivalently, leading to $e^{-\hat{Q}^\text{T}G_0\hat{Q}/2}=\hat{U}_S e^{-\hat{Q}^\text{T}[\oplus_{j=1}^n g_j\mathbb{1}_2]\hat{Q}/2}\hat{U}_S^\dagger$, where $\hat{U}_S\hat{Q}\hat{U}_S^\dagger=S^{-1}\hat{Q}$.
As a consequence,
\begin{align*}
\hat{M}\propto \hat{U}_S e^{v_0^\text{T}[\oplus_{j=1}^n \tanh(g_j/2)\mathbb{1}_2]\hat{Q}}\hat{U}_S^\dagger = e^{v_0^\text{T}[\oplus_{j=1}^n \tanh(g_j/2)\mathbb{1}_2]S^{-1}\hat{Q}},
\end{align*}
where we have used Eq.~(4).

When $G_{\text{M}}\neq0$, the operator $\hat{M}$ can always be written in the Gibbs-exponential form,
\begin{align}\label{final}
\hat{M}\propto e^{-(\hat{Q}-u_\text{M})^{\text{T}} G_{\text{M}}(\hat{Q}-u_\text{M})/2}.
\end{align}
where $u_\text{M}=(e^{-i\Omega G_{\text{M}}}-1)^{-1}m$.
Therefore, $\hat{M}$ can be written as
\begin{align*}
\hat{M}\propto\exp\left[-\frac{1}{2}\hat{Q}^{\text{T}} G_{\text{M}}\hat{Q}- v_\text{M}^{\text{T}}\hat Q\right].
\end{align*}
Here, $v_\text{M} =0$ if $v_0=0$, $v_\text{M}=G_{\text{M}}u_\text{M}$ if $G_0\neq G_1$, and $G_{\text{M}}=0$ and $v_\text{M}=(S^{-1})^\text{T}[\oplus_{j=1}^n \tanh(g_j/2)\mathbb{1}_2]v_0$ if $G_0=G_1$.
From Eqs.~\eqref{Gk} and \eqref{Gm}, it is clear that $G_{\text{M}}$ is the solution of
\begin{equation*}
	e^{i\Omega G_{\text{M}}} e^{i\Omega G_1} e^{i\Omega G_{\text{M}}} = e^{i\Omega G_0},
\end{equation*}
and the vector $u_\text{M}$ is written as
\begin{align*}
u_\text{M}=&(e^{-i\Omega G_{\text{M}}}-1)^{-1}e^{i\Omega G_1/2}(e^{-i\Omega G_K/2}-1)(e^{-i\Omega G_K}-1)^{-1} \\ \times
&e^{-i\Omega G_1/2}(e^{-i\Omega G_0}-1)v_0.
\end{align*}

Finally, in order to return to the original problem between two general Gaussian states, $\hat{\rho}_0=e^{-\frac{(\hat{Q}-u_0)^{\text{T}}  G_0(\hat{Q}-u_0)}{2}}$ and $\hat{\rho}_1=e^{-\frac{(\hat{Q}-u_1)^\text{T} G_1 (\hat{Q}-u_1)}{2}}$,
we simply introduce a displacement operator $\hat{D}(u_1)$ with $u_0-u_1=v_0$, so that, by using Eq.~(4), we obtain $\hat{M}$ of the original problem written as,
\begin{align}
\hat{M}\propto\hat{D}(u_1)\exp\left[-\frac{1}{2}\hat{Q}^{\text{T}} G_{\text{M}}\hat{Q}- v_\text{M}^{\text{T}}\hat Q\right]\hat{D}^\dagger(u_1).
\end{align}

\subsection{Full equation for $d_1$ and $d_2$.}
\setcounter{equation}{0}
\renewcommand{\theequation}{B\arabic{equation}}
We simplify Eq.~(7) for the single-mode case by assuming $G_0$ and $G_1$ to be Gibbs matrices of a general single-mode Gaussian state and a thermal state, respectively. Expanding the matrices by Pauli matrices and using
\begin{align*}
\cosh{g_1}&=\frac{2\bar{n}_1+1}{2\bar{n}_1(\bar{n}_1+1)}, 
\quad
\sinh{g_1}=\frac{2\bar{n}_1^2+2\bar{n}_1+1}{2\bar{n}_1(\bar{n}_1+1)},
\end{align*}
the left hand side of Eq.~(7) is written as
\begin{align*}
L_0\mathbb{1}_{2}+L_1\hat{\sigma}_x+L_2\hat{\sigma}_y,
\end{align*}
where
\begin{widetext}
\begin{align}
L_0&= (d_1+d_2)\frac{2\bar{n}_1+1}{2\bar{n}_1(\bar{n}_1+1)}\frac{\sinh 2 \sqrt{d_1d_2}}{2\sqrt{d_1 d_2}}+\frac{2 \bar{n}_1^2+2 \bar{n}_1+1}{2\bar{n}_1 (\bar{n}_1+1)}\cosh 2 \sqrt{d_1d_2}, \label{lhs0}\\
L_1&= -i \left(d_1-d_2\right)\bigg(\frac{2 \bar{n}_1^2+2 \bar{n}_1+1}{2\bar{n}_1 (\bar{n}_1+1)}\frac{\sinh 2 \sqrt{d_1d_2}}{2\sqrt{d_1d_2} }+\frac{2\bar{n}_1+1}{2\bar{n}_1(\bar{n}_1+1)}\frac{(d_1 +d_2) \sinh ^2\sqrt{d_1d_2}}{2d_1 d_2}\bigg), \label{lhs1}\\
L_2&=\frac{2\bar{n}_1^2+2\bar{n}_1+1}{2\bar{n}_1(\bar{n}_1+1)}\frac{\left(d_1-d_2\right)^2 -\left(d_1+d_2\right){}^2 \cosh 2 \sqrt{d_1d_2}}{4 d_1 d_2}-\frac{2\bar{n}_1+1}{2\bar{n}_1(\bar{n}_1+1)}\frac{2 \sqrt{d_1d_2} \left(d_1+d_2\right) \sinh 2 \sqrt{d_1d_2}}{4 d_1 d_2}.\label{lhs2}
\end{align}
\end{widetext}
The right-hand side, on the other hand, is written as
\begin{align*}
R_0\mathbb{1}_{2}+R_1\hat{\sigma}_x+R_2\hat{\sigma}_y,
\end{align*}
\clearpage
\noindent where
\begin{align}
R_0&=\frac{2 \bar{n}_0^2+2 \bar{n}_0+1}{2\bar{n}_0 (\bar{n}_0+1)}, \label{rhs0}\\
R_1&=\frac{i (2 \bar{n}_0+1) \sinh 2 r_0}{2\bar{n}_0 (\bar{n}_0+1)}, \label{rhs1}\\
R_2&=-\frac{(2 \bar{n}_0+1) \cosh 2 r_0}{2\bar{n}_0 (\bar{n}_0+1)}. \label{rhs2}
\end{align}
Equations of B1 to B6 enable $d_1$ and $d_2$ to be written as functions of $r_0, \bar{n}_0$, and $\bar{n}_1$.

\subsection{Pure state limit}
\setcounter{equation}{0}
\renewcommand{\theequation}{C\arabic{equation}}
Consider a single-mode state with a diagonal covariance matrix of
\begin{equation*}
	 V = \begin{pmatrix}
		 \frac12 + \epsilon & 0 \\ 0 & \frac12 + \epsilon 
	 \end{pmatrix}~. 
\end{equation*}
Such state is pure in the limit of $\epsilon\to 0$. The analysis can be trivially extended to a non-diagonal 
case by adding a squeezing operation $SVS^T$. One can find that 
\begin{align}
	e^{i\Omega G} &= \frac{W-\openone}{W+\openone} = \left(\frac1\epsilon +1\right) P + \epsilon Q +  O(\epsilon^2)~,
	\label{GP}
	\\
	e^{-i\Omega G} &= \frac{W+\openone}{W-\openone} = \left(\frac1\epsilon+1\right)  Q + \epsilon P +  O(\epsilon^2)~,
	\label{GQ}
\end{align}
where $W=-2V i\Omega$ and 
\begin{align*}
	P = \frac12 \begin{pmatrix}
		1 & -i \\ i & 1
	\end{pmatrix}, \quad
	Q = \openone - P.
\end{align*}
Note $P^2= P$ and $Q^2=Q$, so they are projection operators. The Gibbs matrix of the operator $\hat{M}$ satisfies 
\begin{equation} 
		e^{ i\Omega G_1} 
		= 	e^{-i\Omega G_{\text{M}}}e^{ i\Omega G_0}	e^{-i\Omega G_{\text{M}}}  ~.
		\label{GM}
\end{equation}
In the limit where $G_1$ corresponds to the pure state $\ket{\psi_1}\bra{\psi_1}$, we use Eqs.~\eqref{GP} to write $e^{ i\Omega G_1}  \approx \frac{P}{\epsilon}$.
Then a possible solution for $ e^{-i\Omega G_{\text{M}}} $ is $ e^{-i\Omega G_{\text{M}}} \approx \alpha P$ because the above equation \eqref{GM} becomes $\alpha^2 P e^{ i\Omega G_0} P = e^{ i\Omega G_1}  \approx \frac{P}{\epsilon}$, which is approximately true for some $\alpha$. Indeed, for any state $\hat{\rho}_0$ with nonzero overlap with $\hat{\rho}_1$, it is $P e^{ i\Omega G_1} P \propto P$. Therefore, $e^{-i\Omega G_{\text{M}}} \propto P \propto e^{i\Omega G_1}$, namely, $\hat{M} \propto \mathbb{1}-\ket{\psi_1}\bra{\psi_1}$, where all approximations made in the above equations refer to the corrections that disappear in the limit of $\epsilon\to 0$. The operator $\hat{M}$ implies that the measurement with projectors $\{\ket{\psi_1}\bra{\psi_1}, \mathbb{1}-\ket{\psi_1}\bra{\psi_1}\}$ is optimal.
\subsection{The relation between optimal measurements for quantum fidelity and quantum Fisher information}
\setcounter{equation}{0}
\renewcommand{\theequation}{D\arabic{equation}}

Let $\hat{\rho}_0=\hat{\rho}+d\hat{\rho}$ and $\hat{\rho}_1=\hat{\rho}$.
For simplicity, we assume $\hat{\rho}$ is a full-rank state, which implies that $\hat{\rho}_0$ and $\hat{\rho}_1$ are full-rank states.
Let $\sqrt{\hat{\rho}_1^{1/2}\hat{\rho}_0\hat{\rho}_1^{1/2}}=\hat{\rho}+X$, where $X\propto d\hat{\rho}$. Taking the square, we get
\begin{align*}
\hat{\rho}_0^{1/2}\hat{\rho}_1\hat{\rho}_0^{1/2}&=\hat{\rho}^2+\hat{\rho}^{1/2}d\hat{\rho}\hat{\rho}^{1/2} =\hat{\rho}^2+\hat{\rho} X+X\hat{\rho},
\end{align*}
leading to $\hat{\rho}^{1/2}d\hat{\rho}\hat{\rho}^{1/2}=\hat{\rho} X+X\hat{\rho}$.
For $\hat{\rho}=\sum_kp_k|k\rangle\langle k|$ with $\langle k|l \rangle=\delta_{kl}$, one can show 
\begin{align*}
X_{nm}&=\frac{\sqrt{p_n}\sqrt{p_m}}{p_n+p_m}d\hat{\rho}_{nm}.
\end{align*}
When the states are full rank, the first optimality condition becomes $E_{x}^{1/2}(1-\mu_x\hat{\rho}_1^{-1/2}\sqrt{\hat{\rho}_1^{1/2}\hat{\rho}_0\hat{\rho}_1^{1/2}}\hat{\rho}_1^{-1/2})=0$.
In the limit of small $d\hat{\rho}$,
\begin{align*}
\hat{\rho}_1^{-1/2}\sqrt{\hat{\rho}_1^{1/2}\hat{\rho}_0\hat{\rho}_1^{1/2}}\hat{\rho}_1^{-1/2}&=1+\hat{\rho}^{-1/2}X\hat{\rho}^{-1/2}\\
&=1+\sum_{n,m}\frac{d\hat{\rho}_{nm}}{p_n+p_m}|n\rangle\langle m|=1+\hat{L}d\theta/2,
\end{align*}
where $\hat{L}_\theta d\theta=2\sum_{n,m}d\hat{\rho}_{nm}/(p_n+p_m)|n\rangle\langle m|$ is the SLD operator, so that the condition becomes
\begin{align*}
&\hat{E}_x^{1/2}(1-\mu_x\hat{\rho}_1^{-1/2}\sqrt{\hat{\rho}_1^{1/2}\hat{\rho}_0\hat{\rho}_1^{1/2}}\hat{\rho}_1^{-1/2})\\
=&\hat{E}_x^{1/2}(1-\mu_x(1+\hat{L}_\theta d\theta/2))=0.
\end{align*}
This results in
\begin{align*}
\hat{E}_x^{1/2}(1-\lambda_x\hat{L}_\theta)=0
\end{align*}
with a constant $\lambda_{x}$, which is equivalent to the optimal condition of Eq.~(12) for quantum Fisher information.

Now, we turn to the second condition. For two quantum states that are infinitesimally close, Eq.~(2) can be simplified as
\begin{align*}
\text{Tr}[U\hat{\rho}_0^{1/2}\hat{E}_x\hat{\rho}_1^{1/2}]
&=\text{Tr}[\sqrt{\hat{\rho}_1^{1/2}\hat{\rho}_0\hat{\rho}_1^{1/2}}\hat{\rho}_1^{-1/2}\hat{E}_x\hat{\rho}_1^{1/2}]
\\&=\text{Tr}[(1+\hat{L}_\theta d\theta/2)\hat{E}_x\hat{\rho}]
\in \mathbb{R}.
\end{align*}
One can immediately see that this is equivalent to Eq.~(13).

\subsection{Limit of $G_{\text{M}}$ matrix}
\setcounter{equation}{0}
\renewcommand{\theequation}{E\arabic{equation}}
Consider the problem of estimating parameter $\theta$. The matrix $G_{\text{M}}$ is given by the solution of
\begin{equation*}
	e^{i\Omega G_{\text{M}}} = 
	e^{-i\Omega G_\theta/2}
	\sqrt{
	e^{i\Omega G_\theta/2}
	e^{i\Omega G_{\theta+d\theta}}
	e^{i\Omega G_\theta/2}
  }
	e^{-i\Omega G_\theta/2}.
\end{equation*}
Since the zeroth order of the two matrices $G_\theta$ and $G_{\theta+d\theta}$ is equal in an infinitesimal limit of $d\theta$, the zeroth order of $G_{\text{M}}$ is zero. Therefore, one can write $i\Omega G_{\text{M}} =  C d\theta$ for some unknown matrix $C$ and, similarly, $i\Omega G_{\theta}= A$ and  $i\Omega G_{\theta+d\theta}= A+Bd\theta$ for some matrices $A$ and $B$. From the above 
equation, it can be shown that $C$ is the solution of 
\begin{align*}
	 e^{A+Bd\theta}  \approx e^{A} + e^{A} Cd\theta 
	 + Cd\theta e^{ A} + O(d\theta)^2.
\end{align*}
Using the notation from Ref. \cite{banchi2015}, one may write $e^{i\Omega G_\theta} =
\frac{W_\theta-\openone}{W_{\theta}+\openone}$ and expand the matrices 
$W_\theta$ as $W_{\theta+d\theta}= W_A+W_Bd\theta$ with $W_{\theta}= W_A$.
Therefore,
\begin{align*}
	e^{A+Bd\theta}=e^{i\Omega G_{\theta+d\theta}}&=
	\openone -2 \frac{\openone}{W_{\theta+d\theta}+\openone}
	\\
	&= e^A + 
	\frac{d\theta}2 (e^A-\openone) W_B (e^A-\openone) + 
	O(d\theta)^2
\end{align*}
and $C$ is the solution of 
\begin{equation*}
	e^{A} C + C e^{ A}  = 
	\frac{1}2 (e^A-\openone) W_B (e^A-\openone)~,
\end{equation*}
or $C$ can be implemented into the discrete Lyapunov equation written as
\begin{equation*}
	C -W_\theta^{-1}CW_\theta^{-1} =   W_\theta^{-1}\frac{\partial W_\theta}{\partial \theta} W_\theta^{-1},
\end{equation*}
for which $(W_\theta+\openone)C(W_\theta-\openone)+ (W_\theta-\openone)C(W_\theta+\openone)= 2  W_B$ is used.
The solution of the Lyapunov equation is
\begin{align*}
C=\sum_{m=0}^{\infty} W_\theta^{-m-1}\frac{\partial W_\theta}{\partial \theta}W_\theta^{-m-1},
\end{align*}
and thus,
\begin{align*}
G_{\text{M}}=i\Omega\sum_{m=0}^{\infty} W_\theta^{-m-1}\frac{\partial W_\theta}{\partial \theta}W_\theta^{-m-1}d\theta.
\end{align*}

Especially when $\partial\bar{n}_j/\partial\theta=0$ and isothermal states, i.e. $\bar{n}_j=\bar{n}$ for all $j$,
\begin{align}
C=\sum_{m=0}^{\infty} (-1)^{m+1} W_\theta^{-2m-2}\frac{\partial W_\theta}{\partial \theta}=-
\frac{1}{2(2\bar{n}^2+2\bar{n}+1)}\frac{\partial W_\theta}{\partial \theta},
\end{align}
where we have used $W_\theta^{2}=(2\bar{n}+1)^{2}\mathbb{1}_{2n}$.
Thus,
\begin{align*}
G_{\text{M}}=-\frac{1}{2\bar{n}^2+2\bar{n}+1}\Omega\frac{\partial V_\theta}{\partial \theta}\Omega d\theta.
\end{align*}
It can also be shown that from the definition of $C$ and $W_\theta=-2V_\theta i\Omega$, $G_{\text{M}}$ is also the solution of 
\begin{equation}\label{GMeq3}
	4V_\theta G_{\text{M}} V_\theta +\Omega G_{\text{M}}\Omega + 2 d\theta \, 
	\frac{\partial V_\theta}{\partial \theta}=0.
\end{equation}
Writing in the basis, in which $V_\theta$ is symplectically diagonalized, one can recover the previous result \cite{jiang2014},
\begin{align}
(G_{\text{M}})_{jk}=\frac{2V_\theta^{\text{s}} \frac{\partial V_\theta^{\text{s}}}{\partial\theta} V_\theta^{\text{s}}-\Omega \frac{\partial V_\theta^{\text{s}}}{\partial\theta} \Omega/2}{\lambda_j^2\lambda_k^2-1}d\theta
\end{align}
where the superscript of $s$ denotes operators being transformed by the symplectic operator $S$, $\lambda_j$'s are the symplectic eigenvalues of $V_\theta$, and $S$ is a symplectic matrix that diagonalizes $V_\theta$.

The vector $u_\text{M}$ for an infinitesimal $d\theta$ is written as
\begin{align*}
u_\text{M}&=(-i\Omega G_{\text{M}})^{-1}e^{i\Omega G_\theta/2}(e^{-i\Omega G_\theta/2}-1)(e^{-2i\Omega G_\theta}-1)^{-1}
\\ &\times e^{-i\Omega G_\theta/2}(e^{-i\Omega G_\theta}-1)\frac{\partial u_\theta}{\partial{\theta}}d\theta =G_{\text{M}}^{-1} V_\theta^{-1} \frac{\partial u_\theta}{\partial{\theta}} d\theta/2,
\end{align*}
where we have used $e^{i\Omega G_\theta}=\frac{W_\theta-\mathbb{1}}{W_\theta+\mathbb{1}}$.
Thus,
\begin{align}
v_\text{M}=G_{\text{M}} u_\text{M}=V_\theta^{-1}\frac{\partial u_\theta}{\partial{\theta}} d\theta/2.
\end{align}

As a final remark, we highlight that Eq.~\eqref{GMeq3} with $G_{\text{M}}$ and $v_\text{M}$ facilitates the derivation of the quantum Fisher information, being made as
\begin{align*}
H(\theta)
&=\text{Tr}[\hat{D}^\dagger(u_\theta)\hat{\rho}_\theta\hat{D}(u_\theta) (\hat{Q}^\text{T} G_{\text{M}} \hat{Q}-2 v_\text{M}^\text{T}\hat{Q}+\nu)^2]/d\theta^2 \nonumber \\
&=\text{Tr}[\hat{\rho}_\theta^0 \big(\hat{Q}^\text{T} G_{\text{M}} \hat{Q})^2+4(v_\text{M}^\text{T} \hat{Q})^2+\nu(\hat{Q}^\text{T} G_{\text{M}} \hat{Q})+\nu^2]/d\theta^2 \nonumber \\
&=-\text{Tr}\bigg[\frac{\partial V_\theta}{\partial \theta} G_{\text{M}}\bigg]/(d\theta)+\frac{\partial u_\theta}{\partial{\theta}} V^{-1}_\theta \frac{\partial u_\theta}{\partial{\theta}},
\end{align*}
where $\hat{\rho}_\theta^0=\hat{D}^\dagger(u_\theta)\hat{\rho}_\theta\hat{D}(u_\theta)$ is a Gaussian state with zero mean and the same covariance matrix as $\hat{\rho}_\theta$. We also have used 
$
\text{Tr}[\hat{\rho}_\theta^0 \hat{Q}_n\hat{Q}_m\hat{Q}_l\hat{Q}_k]= \sum_{(m l k)}\text{Tr}[\hat{\rho}_\theta^0 \hat{Q}_n\hat{Q}_m]\text{Tr}[\hat{\rho}_\theta^0\hat{Q}_l\hat{Q}_k]
$, where $(m l k)$ denotes a cyclic permutation, and $\text{Tr}[\hat{\rho}_\theta^0\hat{Q}_n\hat{Q}_m]=V_{nm}+i\Omega_{nm}/2$~\cite{gardiner2004}.
Note that the method we provide above can be straightforwardly applied to multi-parameter cases so as to derive a  quantum Fisher information matrix.

\end{document}